\documentclass[aps,prb,twocolumn]{revtex4}
\usepackage{amsmath}
\usepackage{graphicx}
\begin{document}

\title{Magnetization Dissipation in the Ferromagnetic Semiconductor (Ga,Mn)As}

\author{Kjetil M. D. Hals and Arne Brataas}
\affiliation{ Department of Physics, Norwegian University of Science and Technology, NO-7491, Trondheim, Norway }

\begin{abstract}
We compute the Gilbert damping in (Ga,Mn)As based on the scattering theory of magnetization relaxation. The disorder scattering is included non-perturbatively. In the clean limit,  spin-pumping from the localized
d-electrons to the itinerant holes dominates the relaxation processes. In the diffusive regime, the breathing Fermi-surface effect is balanced by the effects of interband scattering, which cause the Gilbert damping constant to saturate at around 0.005. In small samples, the system shape induces a large anisotropy in the Gilbert damping.
\end{abstract}
\maketitle

\newcommand{\eq}  {  \! = \!  }
\newcommand{\keq} {\!\! = \!\!}
\newcommand{\kadd}{  \! + \!  }
\newcommand{\ksim}{\! \sim \!}

\section{Introduction}

The magnetization dynamics of a ferromagnet can be described phenomenologically by the Landau-Lifshitz-Gilbert (LLG) equation:~\cite{Ralph:jmmm08,LLG_micro}
\begin{equation}
\frac{1}{\gamma}\frac{d\mathbf{M}}{dt} = 
-\mathbf{M}\times \mathbf{H}_{\text{eff}} + 
\mathbf{M}\times \left[   \frac{\tilde{G}(\mathbf{M})}{\gamma^2 M_s^2 }  \frac{d\mathbf{M}}{dt}  \right]   .
\end{equation} 
Here, $\gamma$ is the gyromagnetic ratio, $\mathbf{H}_{\text{eff}}$ is the effective magnetic field (which is the functional derivative of the free energy $\mathbf{H}_{\text{eff}}= -\delta F \left[  \mathbf{M} \right]/ \delta \mathbf{M}$), $\mathbf{M}$ is the magnetization and $M_s$ is its magnitude.
The Gilbert damping constant $\tilde{G}(\mathbf{M})$ parameterizes the dissipative friction process that drives the magnetization towards an equilibrium state.~\cite{Gilbert2004} In the most general case, $\tilde{G}(\mathbf{M})$ is a symmetric positive definite matrix that depends on the magnetization direction; however, it is often assumed to be independent of $\mathbf{M}$ and proportional to the unit matrix, assumptions which are valid for isotropic systems.  
Gilbert damping is important in magnetization dynamics. It determines the magnitudes of the external magnetic fields~\cite{Bland:book} and the current densities~\cite{Ralph:jmmm08} that are required to reorient the magnetization direction of a ferromagnet. Therefore, a thorough understanding of its properties is essential for modeling ferromagnetic systems.  

The main contribution to the Gilbert damping process in metallic ferromagnets is the generation of electron-hole pairs.~\cite{Kambersky:cjpb76, Ralph:jmmm08, LLG_micro, Gilmore:PRL07} A model that captures this process was developed by Kambersky.~\cite{Kambersky:cjpb76} In this model, the electrons are excited by a time-varying magnetization via electron-magnon coupling. If the ferromagnet is in metallic contact with other materials, the spin-pumping into the adjacent leads provides an additional contribution to the magnetization relaxation.~\cite{Tserkovnyak:PRL02} A general theory that captures both of these effects was recently developed.~\cite{Brataas:PRL08} The model expresses the $\tilde{G}(\mathbf{M})$ tensor in terms of the scattering matrix $S$ of the ferromagnetic system ($\mathbf{m}\equiv \mathbf{M}/M_s$):
\begin{equation}
\tilde{G}_{ij}(\mathbf{m})= \frac{\gamma^2 \hbar}{4\pi}\text{Re}
\left\{ \text{Tr}
\left[
\frac{\partial S}{\partial m_i}   \frac{\partial S^{\dagger} }{\partial m_j }   
\right]   
\right\}  . \label{G_expression}
\end{equation}
The expression is evaluated at the Fermi energy.
Instead of $\tilde{G}(\mathbf{M})$, one often parameterizes the damping by the dimensionless Gilbert damping parameter $\tilde{\alpha}  \equiv \tilde{G}/ \gamma M_s $. Eq.~\eqref{G_expression} allows studying both the effects of the system shape and the disorder dependency of the magnetization damping beyond the relaxation time approximation.~\cite{Starikov:prl10}

In anisotropic systems, the Gilbert damping is expected to be a symmetric tensor with non-vanishing off-diagonal terms. We are interested in how this tensor structure influences the dynamics of the precessing magnetization in (Ga,Mn)As. Therefore, to briefly discuss this issue, let us consider a homogenous ferromagnet in which the magnetization direction $\mathbf{m}= \mathbf{m}_0 + \delta \mathbf{m}$ precesses with a small angle around the equilibrium direction $\mathbf{m}_0$ that points along the external magnetic field $\mathbf{H}_{\text{ext}}$.~\cite{Seib:prb09} For clarity, we neglect the anisotropy in the free energy and choose the coordinate system such that $\mathbf{m}_0 = (0\ 0\ 1)$ and $\delta \mathbf{m} = (m_x\  m_y\ 0)$. For the lowest order of Gilbert damping, the LLG equation can be rewritten as:
$\dot{\mathbf{m}}= -\gamma \mathbf{m}\times\mathbf{H}_{\text{ext}} + \gamma\mathbf{m}\times( \tilde{\alpha}\left[  \mathbf{H}_{\text{ext}} \times\mathbf{m} \right])$, where $\tilde{\alpha}\left[  ... \right]$ is the dimensionless Gilbert damping tensor that acts on the vector $\mathbf{H}_{\text{ext}}\times\mathbf{m}$. Linearizing the LLG equation results in the following set of equations for $m_x$ and $m_y$:
\begin{equation}
\begin{pmatrix}
\dot{m}_x \\
\dot{m}_y
\end{pmatrix}
= -\gamma H_{\text{ext}}
\begin{pmatrix}
\alpha_{yy}^{(0)}  &  (1-\alpha_{xy}^{(0)}) \\
-(1+\alpha_{xy}^{(0)}) & \alpha_{xx}^{(0)}
\end{pmatrix}
\begin{pmatrix}
m_x \\
m_y
\end{pmatrix}    .  \label{LinearizedLLG}
\end{equation}   
Here, $\alpha_{ij}^{(0)}$ are the matrix elements of $\tilde{\alpha}$ when the tensor is evaluated along the equilibrium magnetization direction $\mathbf{m}_0$. For the lowest order of Gilbert damping, the eigenvalues of \eqref{LinearizedLLG} are 
$\lambda_{\pm}= \pm i \gamma H_{\text{ext}} -  \gamma H_{\text{ext}}\alpha  $, and the eigenvectors describe a precessing magnetization with a characteristic life time $\tau= (\alpha \gamma H_{\text{ext}})^{-1}$. The effective damping coefficient $\alpha$ is:~\cite{Seib:prb09}  
\begin{equation}
\alpha \equiv \frac{1}{2}\left(  \alpha_{xx}^{(0)}  +  \alpha_{yy}^{(0)}  \right)  . \label{EffectiveDamping}
\end{equation}
The value of $\alpha$ is generally anisotropic and depends on the static magnetization direction $\mathbf{m}_0$. The magnetization damping is accessible via ferromagnetic resonance (FMR) experiments by measuring the linear relationship between the FMR line width and the precession frequency. This
linear relationship is proportional to $\tau^{-1}$ and thus depends linearly on $\alpha$. Therefore, an FMR experiment can be used to determine the effective damping coefficient $\alpha$. In contrast, the off-diagonal terms, $\alpha_{xy}^{(0)}$ and $\alpha_{yx}^{(0)}$, do not contribute to the lowest order in the damping and are difficult to probe experimentally. 

In this paper, we use Eq.~\eqref{G_expression} to study the anisotropy and disorder dependency of the Gilbert damping in the ferromagnetic semiconductor (Ga,Mn)As. Damping coefficients of this material in the range of $\alpha\sim 0.004-0.04$ for annealed samples have been reported.~\cite{Sinova:prb04,Matsuda:pb08,Wirthmann:apl08,Khazen:prb08} The damping is anisotropically dependent on the magnetization direction.~\cite{Sinova:prb04,Matsuda:pb08,Khazen:prb08} 
The few previous calculations of the Gilbert damping constant in this material have indicated that $\alpha\sim 0.003-0.04$.~\cite{Sinova:prb04,Tserkovnyak:apl04,Garate:prb09_1,Garate:prb09_2} These theoretical works have included the effects of disorder phenomenologically, for instance, by applying the relaxation time approximation.
In contrast, Eq.~\eqref{G_expression} allows for studying the disorder effects fully and non-perturbatively for the first time. In agreement with Ref.~\onlinecite{Tserkovnyak:apl04}, we show that spin-pumping from the localized
d-electrons to the itinerant holes dominates the damping process in the clean limit. In the diffusive regime,  the breathing Fermi-surface effect is balanced by effects of the interband transitions, which cause the damping to saturate.  In determining the anisotropy of the Gilbert damping tensor, we find that the shape of the sample is typically more important than the effects of the strain and the cubic symmetry in the GaAs crystal.~\cite{foot_1}  This shape anisotropy of the Gilbert damping in (Ga,Mn)As has not been reported before and provides a new direction for engineering the magnetization relaxation. 

\section{Model}
The kinetic-exchange effective Hamiltonian approach gives a reasonably good description of the electronic properties of (Ga,Mn)As.~\cite{jungwirth:rmd06} The model assumes that the electronic states near the Fermi energy have the character of the host 
material GaAs and that the spins of the itinerant quasiparticles interact with the localized magnetic Mn impurities (with spin 5/2) via the isotropic Heisenberg exchange interaction. If the s-d exchange interaction is modeled by a mean field,
the effective Hamiltonian takes the form:~\cite{jungwirth:rmd06,abolfath:prb01}
\begin{equation}
H = H_{\text{Holes}} + \mathbf{h}(\mathbf{r})\cdot \mathbf{s}, \label{Hamiltonian_0}
\end{equation} 
where $H_{\text{Holes}}$ is the $\mathbf{k\cdot p}$ Kohn-Luttinger Hamiltonian describing the valence band structure of GaAs and $\mathbf{h}(\mathbf{r})\cdot \mathbf{s}$ is a mean field description of the s-d exchange interaction between the itinerant holes and the local magnetic impurities ($\mathbf{s}$ is the spin operator). The exchange field $\mathbf{h}$ is antiparallel
to the magnetization direction $\mathbf{m}$. The explicit form of $H_{\text{Holes}}$ that is needed for realistic modeling of the band structure of GaAs depends on the doping level of the system. Higher doping levels often require an eight-band model, but  a six- or four-band model may be sufficient for lower doping levels. In the four-band model, the Hamiltonian is projected onto the subspace spanned by the four 3/2 spin states at the top of the GaAs valence band. The six-band model also includes the spin-orbit split-off bands with spin 1/2. The spin-orbit splitting of the spin 3/2 and 1/2 states in GaAs is  341 meV.~\cite{Cardona:book}  We consider a system with a Fermi level of 77 meV when measured from the lowest subband. In this limit, the following four-band model gives a sufficient description: 
\begin{eqnarray}
H &=& \frac{1}{2m} \left[ (\gamma_1 + \frac{5}{2}\gamma_2) \mathbf{p}^2 - 2\gamma_3 (\mathbf{p\cdot J})^2 + \mathbf{h\cdot J}   \right]   +  \nonumber \\
& &\frac{\gamma_3-\gamma_2}{m} (p_x^2J_x^2 + c.p.)  + H_{\text{strain}} + V(\mathbf{r}). \label{Hamiltonian}
\end{eqnarray}
Here, $\mathbf{p}$ is the momentum operator, $J_i$ are the spin 3/2 matrices~\cite{comment2} and $\gamma_1$, $\gamma_2$ and $\gamma_3$ are the Kohn-Luttinger parameters. $V(\mathbf{r})= \sum_i V_i \delta (\mathbf{r}-\mathbf{R}_i)$ is the impurity scattering potential, where $\mathbf{R}_i$ is the position of the impurity $i$ and $V_i$ are the scattering strengths of the impurities~\cite{comment1} that are randomly and
uniformly distributed in the interval $[-V_{0}/2,V_{0}/2]$. $H_{\text{strain}}$ is a strain Hamiltonian and arises because the (Ga,Mn)As system is grown on top of a substrate (such as GaAs).~\cite{Chernyshov:np09} The two first terms in Eq.~\eqref{Hamiltonian} have spherical symmetry, and the term proportional to $\gamma_3-\gamma_2$ represents the effects of the cubic symmetry of the GaAs crystal. Both this cubic symmetry term~\cite{Baldereschi:prb73} and the strain Hamiltonian~\cite{Chernyshov:np09} are small compared to the spherical portion of the Hamiltonian. A numerical calculation shows that they give a correction to the Gilbert damping on the order of 10\%. However, the uncertainty of the numerical results, due to issues such as the sample-to-sample disorder fluctuations, is also about 10\%; therefore, we cannot conclude how these terms influence the anisotropy of the Gilbert damping.  Instead, we demonstrate that the shape of the system is the dominant factor influencing the anisotropy of the damping. Therefore, we disregard the strain Hamiltonian $H_{\text{strain}}$ and the term proportional to $\gamma_3-\gamma_2$ in our investigation of the Gilbert damping.     
 \begin{figure}[ht] 
\centering 
\includegraphics[scale=1.0]{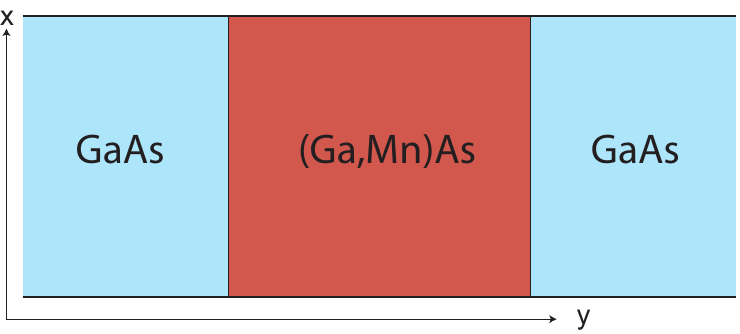} 
\caption{We consider a (Ga,Mn)As system attached along the $\left[ 0 1 0 \right]$ direction to infinite ballistic GaAs leads. The scattering matrix is calculated for the (Ga,Mn)As layer and one 
lattice point into each of the leads. The magnetization is assumed to be homogenous. In this paper, we denote the $\left[ 1 0 0 \right]$ direction as the x-axis, the $\left[ 0 1 0 \right]$ direction as the y-axis and
the $\left[ 0 0 1 \right]$ direction as the z-axis  }
\label{Fig1} 
\end{figure} 

We consider a discrete (Ga,Mn)As system with transverse dimensions $L_{x}%
\in \left\{ 17, 19, 21  \right\}$~nm, $L_{z}\in \left\{11, 15, 17\right\}$~nm and $L_{y}= 50$~nm and
connected to infinite ballistic GaAs leads, as illustrated in Fig.~\ref{Fig1}.  The leads are modeled as being identical to the (Ga,Mn)As system, except for the magnetization and disorder.
The lattice constant is $1$~nm, which is much less than the Fermi wavelength $\lambda_{F}\sim10$~nm. The Fermi energy is 0.077~eV when measured from the lowest subband edge. 
The Kohn-Luttinger parameters are 
$\gamma_{1}=7.0$ and $\gamma_{2}=\gamma_3=2.5$, implying that we apply the spherical approximation for 
the Luttinger Hamiltonian, as mentioned above.~\cite{Baldereschi:prb73}
We use  $|\mathbf{h}|=0.032$~eV for the exchange-field strength.  
To estimate a typical saturation value of the magnetization, we use $M_{s}
=10|\gamma|\hbar x/a_{\mathrm{GaAs}}^{3}$ with $x=0.05$ as
the doping level and $a_{\mathrm{GaAs}}$ as the lattice constant for GaAs.~\cite{comment3}

The mean free path $l$ for the impurity strength $V_{0}$ is calculated by fitting
the average transmission probability $T=\left\langle G\right\rangle /G_{sh}$
to $T(L_{y})=l/(l+L_{y})$,~\cite{Datta:book} where $G_{sh}$ is the Sharvin
conductance and $\left\langle G\right\rangle $ is the conductance for a system
of length $L_{y}$. 

The scattering matrix is calculated numerically using a stable transfer matrix
method.~\cite{Usuki:prb95} The disorder effects are fully and
non-perturbatively included by the ensemble average $\left\langle  \alpha \right\rangle
=\sum_{n=1}^{N_{I}}  \alpha _{n}/N_{I}$, where $N_{I}$ is the number
of different impurity configurations. All the coefficients are averaged until an
uncertainty $\delta\left\langle \alpha\right\rangle =\sqrt{\left(
\left\langle \alpha^{2}\right\rangle -\left\langle \alpha \right\rangle
^{2}\right)  /N_{I}}$ of less than $10\%$ is achieved.
The vertex corrections are exactly included in the scattering formalism.

\section{Results and discussion}
\begin{figure}[ht] 
\centering 
\includegraphics[scale=1.0]{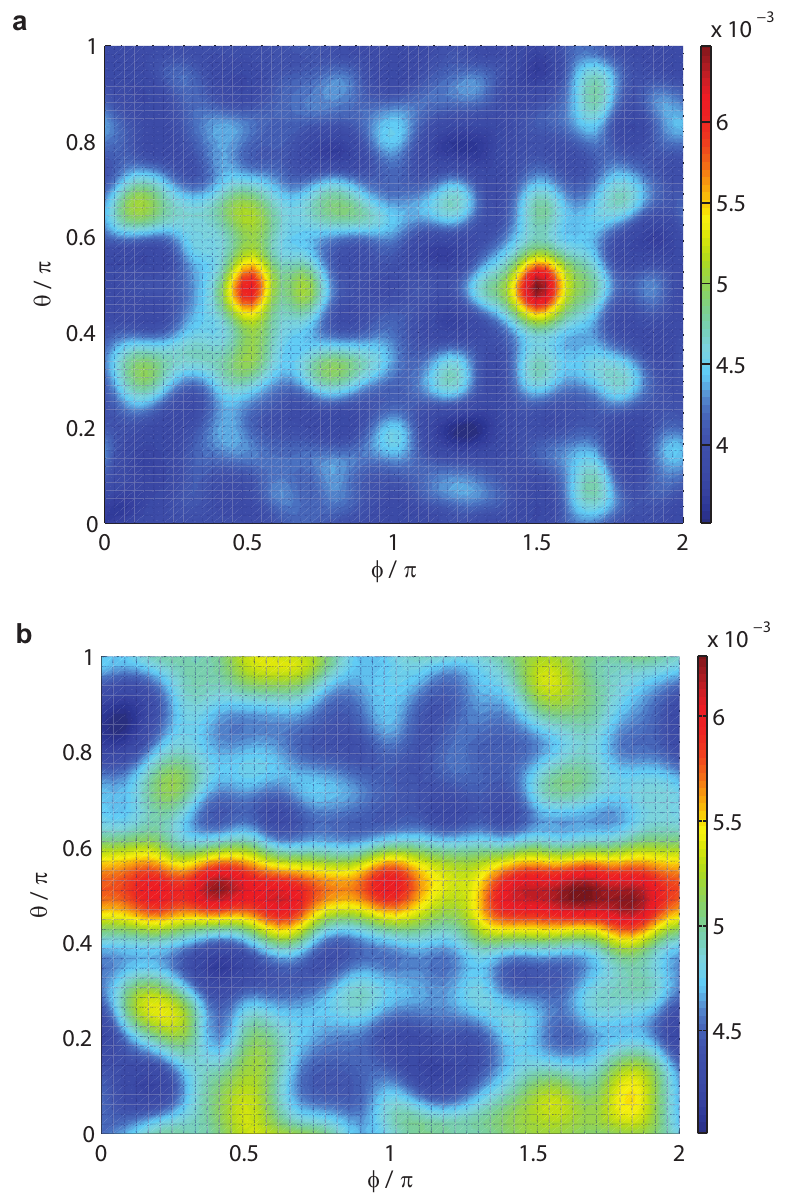} 
\caption{   ({\bf a}) The dimensionless Gilbert damping parameter $\alpha$ as a function of the magnetization direction for a system where $L_x=17$~nm, $L_y=50$~nm and $L_z=17$~nm.    
({\bf b}) The dimensionless Gilbert damping parameter $\alpha$ as a function of the magnetization direction for a system where $L_x=21$~nm, $L_y=50$~nm and $L_z=11$~nm.  
 Here, $\theta$ and $\phi$ are the polar and azimuth angles, respectively, that describe the local magnetization direction $\mathbf{m} = \left( \sin\theta\cos\phi,\: \sin\theta\sin\phi,\: \cos\theta \right) $.
 In both plots, the mean free path is $l\sim 22$~nm}
\label{fig:Fig2} 
\end{figure} 
Without disorder, the Hamiltonian describing our system is rotationally symmetric around the axis parallel to $\mathbf{h}$.
Let us briefly discuss how this influences the particular form of the Gilbert damping tensor $\tilde{\alpha}$.~\cite{Steiauf:prb05} 
For clarity, we choose the coordinate axis such that the exchange field points along the z-axis. In this case, the Hamiltonian 
is invariant under all rotations $R_z$ around the z-axis. This symmetry requires the energy dissipation 
$\dot{E}\propto \dot{\mathbf{m}}^T \tilde{\alpha} \dot{\mathbf{m}}$ ($\dot{\mathbf{m}}^T$ is the transposed of $\dot{\mathbf{m}}$) of the magnetic system  to be invariant under the coordinate transformations
$\mathbf{r}^{'}= R_z \mathbf{r}$ (i.e., $ (\dot{\mathbf{m}}^{'})^T \tilde{\alpha}^{'} \dot{\mathbf{m}}^{'}= \dot{\mathbf{m}}^T \tilde{\alpha} \dot{\mathbf{m}}$ where $\mathbf{m}^{'}= R_z \mathbf{m}$ and
$\tilde{\alpha}^{'}$ is the Gilbert damping tensor in the rotated coordinate system). Because $\tilde{\alpha}$ only depends on the direction of $\mathbf{m}$, which is unchanged under the coordinate transformation
$R_z$,  $\tilde{\alpha}= \tilde{\alpha}^{'}$ and  $  R_z^T \tilde{\alpha} R_z= \tilde{\alpha}$. Thus, $\tilde{\alpha}$ and $R_z$ have common eigenvectors $\left\{ \left|  \pm\right\rangle \equiv (\left|  x\right\rangle \pm i\left|  y\right\rangle)/\sqrt{2},\:  \left|  z\right\rangle \right\} $, and the spectral decomposition of $\tilde{\alpha}$ is
$\tilde{\alpha}= \alpha_{+} \left|  + \right\rangle \left\langle + \right| \   \  + \  \  \alpha_{-} \left|  - \right\rangle \left\langle - \right| \   \  + \  \  \alpha_{z} \left|  z \right\rangle \left\langle z \right|$. Representing the damping tensor in the  
$\left\{ \left|  x\right\rangle ,\: \left|  y\right\rangle,\:  \left|  z\right\rangle \right\} $ coordinate basis yields  $\alpha\equiv \tilde{\alpha}_{xx} = \tilde{\alpha}_{yy} = (\alpha_{+} + \alpha_{-})/2 $, $\tilde{\alpha}_{zz}= \alpha_z$, 
$\tilde{\alpha}_{yx}=\tilde{\alpha}_{xy}=0$,  and $\alpha_{+}=\alpha_{-}$. The last equality results from real tensor coefficients.  However, $\alpha_{zz}$ cannot be determined uniquely from the 
energy dissipation formula $\dot{E}\propto \dot{\mathbf{m}}^T \tilde{\alpha} \dot{\mathbf{m}}$ because $\dot{\mathbf{m}}$ is perpendicular to the z-axis. Therefore, $\alpha_{zz}$ has no physical significance and the energy dissipation is 
governed by the single parameter $\alpha$. For an infinite system, this damping parameter does not depend on the specific direction of the magnetization, i.e., it is isotropic  
because the symmetry of the Hamiltonian is not directly linked to the crystallographic axes of the underlying crystal lattice (when the cubic symmetry term in Eq.~\eqref{Hamiltonian} is disregarded). For a finite system,
the shape of the system induces anisotropy in the magnetization damping. This effect is illustrated in Fig.~\ref{fig:Fig2}, which plots the effective damping in Eq.~\eqref{EffectiveDamping} as a function of the magnetization directions 
for different system shapes. When the cross-section of the conductor is deformed from a regular shape to the shape of a thinner system, the anisotropy of the damping changes.  The magnetization damping varies from a minimum value of around 0.004 to a maximum value of 0.006, e.g., the anisotropy is around 50\%. The relaxation process is largest along the axis where the ballistic leads are connected, i.e., the y-axis. This shape anisotropy is about four- to five-times stronger than the anisotropy induced by the strain and the cubic symmetry terms in the Hamiltonian \eqref{Hamiltonian}, which give corrections of about 10 percent.  For larger systems, we expect this shape effect to become less dominant. In these systems, the anisotropy of the bulk damping parameter, which is 
induced by the anisotropic terms in the Hamiltonian, should play a more significant role. The determination of the system size when the strain and cubic anisotropy become comparable to the shape anisotropy effects is beyond the scope of this paper because the system size is restricted by the computing time. However, this question should be possible to investigate experimentally by measuring the anisotropy of the Gilbert damping as a function of the film thickness.

\begin{figure}[ht] 
\centering 
\includegraphics[scale=1.0]{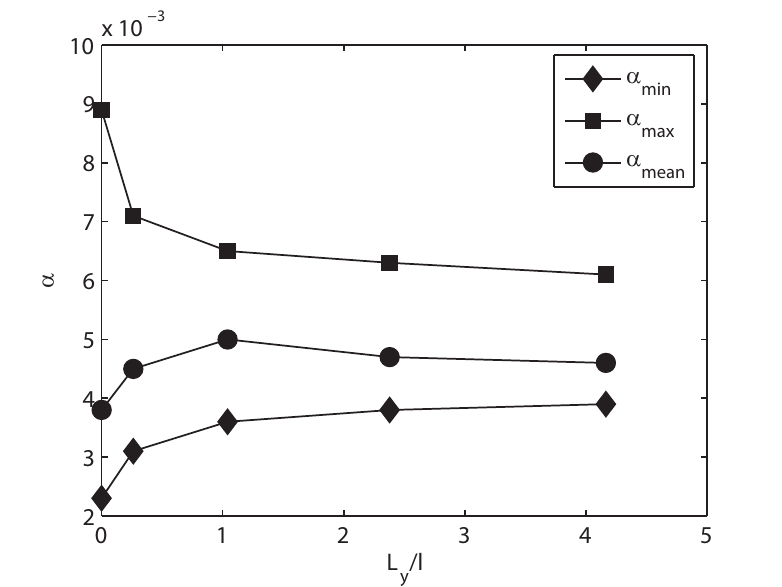} 
\caption{The effective dimensionless Gilbert damping \eqref{EffectiveDamping} as a function of the disorder. Here, $l$ is the mean free path and $L_y$ is the length of the ferromagnetic system in the transport direction. 
$\alpha_{\text{min}}$ and $\alpha_{\text{max} }$ are the minimum and maximum values of the anisotropic Gilbert damping parameter and $\alpha_{\text{mean}}$ is the effective damping parameter averaged over all the magnetization directions. The system dimensions are $L_x=19$~nm, $L_y=50$~nm and $L_z=15$~nm.  }
\label{fig:Fig3} 
\end{figure} 
We next investigate how the magnetization relaxation process depends on the disorder. 
Ref.~\onlinecite{Tserkovnyak:apl04} derives an expression that relates the Gilbert damping parameter to the spin-flip rate $T_2$ of the system: $\alpha \propto T_2 (  1 +  (T_2)^2  )^{-1}$.
In the low spin-flip rate regime, this expression  scales with $T_2$ as $\alpha\propto T_2^{-1}$, while the damping parameter is proportional to $\alpha\propto T_2$ in the opposite limit .
As explained in  Ref.~\onlinecite{Tserkovnyak:apl04}, the low spin-flip regime is dominated by the spin-pumping process in which angular momentum is transferred to the itinerant particles; the transferred spin is then relaxed with a rate 
proportional to $T_2^{-1}$. This process appears inside the ferromagnet itself, i.e., the spin is transferred from the magnetic system to the itinerant particles in the ferromagnet, which are then relaxed {\it within} the ferromagnet. Therefore, this relaxation mechanism is a bulk process and should not be confused with the spin-pumping interface effect across the normal metal $|$ ferromagnet interfaces reported in Ref.~\onlinecite{Tserkovnyak:PRL02}.  In (Ga, Mn)As, this bulk process corresponds to spin-pumping from the d-electrons of the magnetic Mn impurities to the itinerant spin 3/2 holes in the valence band of the host compound GaAs. The transfer of spin to the holes is then relaxed by the impurity scattering within the ferromagnet. By contrast, the opposite limit is dominated by the breathing Fermi-surface mechanism. In this mechanism, the spins of the itinerant particles are not able to follow the local magnetization direction adiabatically and lag behind with a delay time of $T_2$. In our system, which has a large spin-orbit coupling in the band structure, we expect the spin-flip rate to be proportional to the mean free path ($l \propto T_2$).~\cite {Hilton:prl02} 
The effective dimensionless Gilbert damping \eqref{EffectiveDamping} is plotted as a function of disorder in Fig.~\ref{fig:Fig3}. The damping ($\alpha_{\text{mean}}$) partly shows the same behavior as that reported
in Ref.~\onlinecite{Tserkovnyak:apl04}. For clean systems (i.e.,  those with a low spin-flip rate regime), the damping increases with disorder. In such a regime, the transfer of angular momentum to the spin 3/2 holes is the dominant damping process, i.e., the bulk spin-pumping process dominates.  $\alpha_{\text{mean}}$ starts to decrease for smaller mean free paths, implying that the main contribution to the damping comes from the breathing Fermi-surface process. Refs.~\onlinecite{Sinova:prb04,Garate:prb09_1,Garate:prb09_2} have reported that the Gilbert damping may start to increase as a function of disorder in dirtier samples. The interband transitions become more important with decreasing quasi-particle life times and start to dominate the intraband transitions (The intraband transitions give rise to the breathing Fermi-surface effect).    
We do not observe an increasing behavior in the more diffusive regime, but we find that the damping saturates at a value of around 0.0046 (See Fig.~\ref{fig:Fig3}). In this regime, we believe that the breathing Fermi-surface effect is balanced by the interband transitions. The damping does not vanish in the limit $1/l=0$ due to scattering at the interface between the GaAs and (Ga,Mn)As layers in addition to spin-pumping into the adjacent leads (an interface spin-pumping effect, as explained above). Fig.~\ref{fig:Fig3} shows that the shape anisotropy of the damping is  reduced by disorder because the difference between the maximum ($\alpha_{\text{max}}$) and minimum ($\alpha_{\text{min}}$) values of the damping parameter decrease with disorder. We anticipate this result because disorder increases the bulk damping effect, which is expected to be isotropic for an infinite system.
 
\section{Summary}
In this paper, we studied the magnetization damping in the ferromagnetic semiconductor (Ga,Mn)As. The Gilbert damping was calculated numerically using a recently developed scattering matrix theory of magnetization dissipation.\cite{Brataas:PRL08} We conducted a detailed non-perturbative study of the effects of disorder and an investigation of the damping anisotropy induced by the shape of the sample.

Our analysis showed that the damping process is mainly governed by three relaxation mechanisms. In the clean limit with little disorder, we found that the magnetization dissipation is dominated by spin-pumping from the d-electrons to
the itinerant holes. For shorter mean free paths, the breathing Fermi-surface effect starts to dominate, which causes the damping to decrease. In the diffusive regime, the breathing Fermi-surface effect is balanced by the interband transitions and the effective damping parameter saturates at a value on the order of 0.005. 

For the small samples considered in this study, we found that the shape of the system was typically more important than the anisotropic terms in the Hamiltonian for the directional dependency of the damping parameter.  This shape anisotropy has not been reported before and offers a new way of manipulating the magnetization damping. 

\section{Acknowledgments}

This work was partially supported by the European Union FP7 Grant No.~251759 ``MACALO".


\begin{thebibliography}{99}
\bibitem{foot_1}This shape anisotropy in the Gilbert damping should not be confused with the shape anisotropy (in the anisotropy field) caused by surface dipoles in non-spherical systems.
\bibitem{comment2}Note that in Eq.~\eqref{Hamiltonian} the spin operator $\mathbf{s}$ (in the p-d exchange term) is represented in the basis consisting of the four spin 3/2 states ($\mathbf{s}= \mathbf{J}/3$).
The factor $1/3$ is absorbed in the exchange field $\mathbf{h}$.
\bibitem{comment1}In the discrete version of Eq.~\eqref{Hamiltonian}, as used in the numerical calculation, we have one impurity at each lattice site.
\bibitem{comment3}The prefactor of 10 comes from 4 Ga atoms per unit cell times spin 5/2 per substitutional Mn, which are assumed to be fully polarized. The reduction of the net magnetization due to the interstitial Mn ions and p holes are disregarded in our estimate.
\end{thebibliography}
\end{document}